\documentclass[lettersize,journal]{IEEEtran}
\usepackage{amsmath,amsfonts}
\usepackage{xcolor}
\usepackage{algorithmic}
\usepackage{algorithm}
\usepackage{array}
\usepackage[caption=false,font=normalsize,labelfont=sf,textfont=sf]{subfig}
\usepackage{textcomp}
\usepackage{stfloats}
\usepackage{url}
\usepackage{verbatim}
\usepackage{graphicx}
\usepackage{cite}
\usepackage{mathtools}
\usepackage{bm,amsmath, amssymb}
\usepackage{amsmath}
\usepackage{amsfonts}
\usepackage {amssymb,amsmath}

\newcommand {\Define} {\stackrel {\Delta} {=}  }
\newcommand{\mya}{\mathrel{\overset{\makebox[0pt]{{\tiny(a)}}}{=}}}
\newcommand{\myb}{\mathrel{\overset{\makebox[0pt]{{\tiny(b)}}}{=}}}
\newcommand{\myc}{\mathrel{\overset{\makebox[0pt]{{\tiny(c)}}}{=}}}
\newcommand{\myd}{\mathrel{\overset{\makebox[0pt]{{\tiny(d)}}}{=}}}

\newtheorem{theorem}{Theorem}

\begin{document}

\title{Low-Complexity Frequency Domain Equalization of Zak-OTFS in Doubly-Spread Channels}

\author{\small Saif Khan Mohammed, Sandesh Rao Mattu$^*$, Nishant Mehrotra$^*$, Venkatesh Khammammetti and Robert Calderbank
        % <-this % stops a space
\thanks{S. K. Mohammed is with Department of Electrical Engineering, Indian Institute of Technology Delhi, India (E-mail: saifkmohammed@gmail.com).
%S. K. Mohammed is also associated with Bharti School of Telecom. Technology and Management (BSTTM), IIT Delhi.
Sandesh, Nishant, Venkatesh, and Robert are with the Department of Electrical and Computer Engineering, Duke University, Durham, NC, 27708, USA (email: \{sandesh.mattu,~nishant.mehrotra,~venkatesh.khammammetti,~robert.calderbank\}@duke.edu).\\
$^*$ denotes equal contribution.}

}% <-this % stops a space
%\thanks{Manuscript received April 19, 2021; revised August 16, 2021.}}



\maketitle

\vspace{-6mm}

\begin{abstract}
We communicate over wireless channels by first estimating and then equalizing the effective channel. In Zak-OTFS (orthogonal time frequency space) modulation the carrier waveform is a pulse in the delay-Doppler (DD) domain, formally a quasi-periodic localized function with specific periods along delay and Doppler. When the channel delay spread is less than the delay period, and the channel Doppler spread is less than the Doppler period, the response to a single Zak-OTFS carrier provides an image of the scattering environment and can be used to predict the effective channel at all other carriers. This makes DD domain channel estimation straightforward, and there is no loss in spectral efficiency since it is possible to design data and pilot signals that are mutually unbiased. However, equalization in the DD domain has high complexity ${\mathcal O}(M^3N^3)$ where $M$, $N$ are respectively the number of delay and Doppler bins in an OTFS frame, and $MN$ is the number of information symbols.

We demonstrate that equalization in the frequency domain (FD) reduces complexity to only
${\mathcal O}(M^2 N^2)$ by taking advantage of the banded structure of the effective FD channel. We also derive a low-complexity method to reconstruct the effective FD channel from the estimated DD domain effective channel.
\end{abstract}

\begin{IEEEkeywords}
Zak-OTFS, DD domain, Low-complexity equalization.
\end{IEEEkeywords}

\section{Introduction}
The performance of CP-OFDM is known to degrade severely in doubly-spread channels due to 
ICI and frequency selective fading \cite{Nee2000, Wang2006}. With $K$ subcarriers, the complete frequency domain (FD) input-output (I/O) relation is characterized by $K^2$ complex channel gains, one between each pair of subcarriers. OFDM is configured to prevent ICI since it is challenging to estimate $K^2$ unknowns even if
pilots are transmitted on each subcarrier. Recently Zak-OTFS modulation has been proposed, which is parameterized by a delay and Doppler period, with information symbols carried by quasi-periodic pulses in the DD domain with period $M$ and $N$ along the delay and Doppler axis, respectively \cite{zakotfs1, zakotfs2, otfsbook}. By choosing the delay and Doppler period to be greater than the delay and Doppler spread of the channel, the complete I/O relation between the transmitted and the received DD domain carriers can be acquired accurately by using a dedicated DD carrier as a pilot, or by spreading the pilot carrier over all DD carriers to avoid any pilot overhead (see \cite{zakotfs1, zakotfs2, spreadpaper}).
In Zak-OTFS the pilot overhead is significantly smaller than that in CP-OFDM since a single DD pilot carrier is sufficient to acquire the complete I/O relation. However, since the DD carriers interfere with each other, joint equalization of all $MN$ carriers must be performed at the receiver. The complexity is high (${\mathcal O}(M^3 N^3)$) since the $MN \times MN$ DD domain effective channel matrix lacks structure \cite{otfsbook}.

In Zak-OTFS, information symbols and pilot are carried by DD domain carriers.
In this paper, we propose to equalize in the FD since the effective FD channel matrix has a banded structure which reduces the complexity of equalization to ${\mathcal O}(M^2 N^2)$ (Section \ref{seczakofdm}).
Section \ref{sec4} proposes a low-complexity method that converts the acquired DD domain effective channel to the FD effective channel which is then used for FD equalization.
%To get an estimate of the effective FD channel matrix (which captures the complete FD I/O relation between all subcarriers), we propose a novel method in Section \ref{sec4} to convert the estimated DD domain I/O relation (acquired from the received pilots in DD domain) to the FD I/O relation at low complexity. 
Simulation results in Section \ref{simsec} confirm that FD equalization indeed achieves the same error rate performance as DD domain equalization, but at significantly lower complexity.

Note that Zak-OTFS is different from Multi-carrier (MC) OTFS introduced in \cite{Hadani2017}.
 Unlike Zak-OTFS, the MC-OTFS I/O relation is \emph{not predictable} in high Doppler spread scenarios, i.e., channel response to any MC-OTFS carrier cannot be accurately estimated from the response to a particular MC-OTFS carrier \cite{zakotfs2, otfsbook}. Unlike Zak-OTFS, it is challenging to acquire the MC-OTFS I/O relation with low pilot overhead, resulting in degradation in error rate performance with increasing channel Doppler spread \cite{zakotfs2}.

\begin{figure*}[h]%[t]
\vspace{-9mm}
\centering
\includegraphics[width=14.8cm, height=4.9cm]{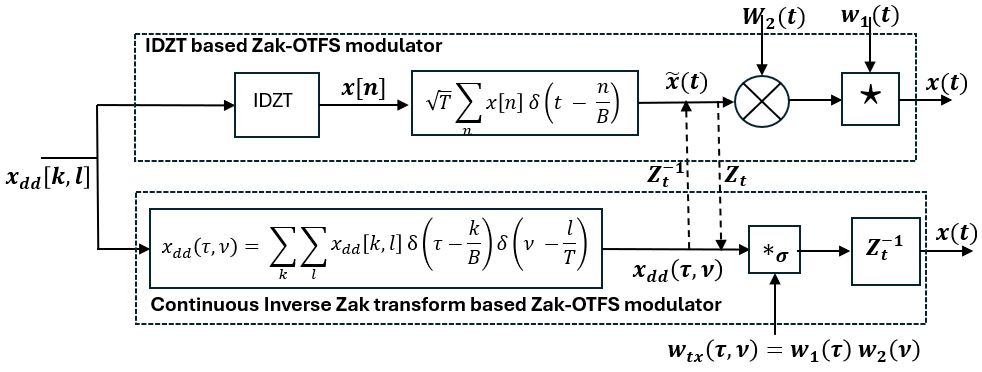}
\vspace{-2mm}
\caption{
Two equivalent implementations of signal processing at the transmitter, (i) inverse Discrete Zak Transform (IDZT) based processing, and (ii) continuous inverse Zak transform (${\mathcal Z}_t^{-1}$) based processing.}
\label{fig1}
%\vspace{-1em}
\vspace{-5mm}
\end{figure*}
\section{System model}
\label{secsysmodel}
We consider a doubly-spread channel with DD spreading function $h_{\mbox{\scriptsize{phy}}}(\tau, \nu)$. The min. and max. delay and Doppler co-ordinates of the support set of $h_{\mbox{\scriptsize{phy}}}(\tau, \nu)$ are denoted by $(0 \,, \, \tau_{max})$ and $(-\nu_{\max}, \nu_{max})$ respectively. Let $x(t)$ denote the transmitted time-domain (TD) signal/frame which has its energy approximately localized to the TD interval $[-(\tau_{\max}+T)/2 \,,\, (\tau_{\max} + T)/2$ and the FD interval $[ -B/2 - \nu_{\max} \,,\, B/2 + \nu_{\max}]$. Here $T$ and $B$ represent the length of the intervals containing the information and the remaining is for cyclic prefix (CP). The received TD signal is given by \cite{Bello}

{\vspace{-4mm}
\small
\begin{eqnarray}
\label{eqnrt2}
    r(t) & \hspace{-3mm} = & \hspace{-3mm}  \iint h_{\mathrm{phy}}(\tau, \nu) x(t - \tau) \, e^{j 2 \pi \nu (t - \tau)} \, d\tau \, d\nu \, + \, n(t),
\end{eqnarray}\normalsize}where $n(t)$ is AWGN with power spectral density $N_0$ Watt/Hz. 
\subsection{Transmitter Signal Processing}
Let $x[k,l]$, $k=0,1,\cdots, M-1$, $l=0,1,\cdots, N-1$ denote the array of $MN = B T$ information symbols in a frame.
At the transmitter, $x[k,l]$ is embedded into a quasi-periodic discrete DD domain signal (TD realization is defined only for quasi-periodic signals \cite{otfsbook})
\begin{eqnarray}
    x_{dd}[k,l] & = & x[k \, \mbox{\small{mod}} \, M, l \, \mbox{\small{mod}} \, N] \, e^{j 2 \pi \lfloor \frac{k}{M} \rfloor \frac{l}{N}}
\end{eqnarray}with periods $M$ and $N$ along the discrete delay and Doppler axis, i.e., for all $k, l, n,m \in {\mathbb Z}$
\begin{eqnarray}
\label{qpeqn234}
    x_{dd}[k + nM, l + mN] & = & e^{j 2 \pi n \frac{l}{N}} \, x_{dd}[k,l].
\end{eqnarray}
%For the OFDM modulator based proposed Zak-OFDM (see Fig.~\ref{fig1}), 

The discrete DD signal is then converted to its discrete-time (DT) realization through the %Inverse Discrete Fourier Transform (IDFT) inside the OFDM modulator (see Fig.~\ref{fig1})
%{\vspace{-4mm}
%\small
%\begin{eqnarray}
%    x[n] & = & \frac{1}{\sqrt{MN}} \sum\limits_{i=0}^{MN -1} S[i] \, e^{j 2 \pi n \frac{i}{MN}}  \,\,,\,\, n \in {\mathbb Z}.
%\end{eqnarray}\normalsize}Note that $x[n]$ is also periodic with period $MN$. The composition of IDFZT followed by IDFT is referred to as the
Inverse Discrete Zak transform (IDZT) (see Chapter $8$ in \cite{otfsbook}) and is given by

{\vspace{-4mm}
\small
\begin{eqnarray}
\label{eqn66n}
x[n] & = & \mbox{\small{IDZT}}\left( x_{dd}[k,l] \right) \, = \, \frac{1}{\sqrt{N}} \sum\limits_{l=0}^{N-1} x_{dd}[k,l],
\end{eqnarray}\normalsize}$n \in {\mathbb Z}$. DT signal $x[n]$ is then converted to

{\vspace{-4mm}
\small
\begin{eqnarray}
{\Tilde x}(t) \Define \sqrt{T} \sum\limits_{n \in {\mathbb Z}} x[n] \delta(t - n/B).
\end{eqnarray}\normalsize}Pulse shaping of ${\Tilde x}(t)$ gives the transmit signal

{\vspace{-4mm}
\small\begin{eqnarray}
\label{pulseshapeeqn}
    x(t) & = & \sqrt{T} w_1(t) \star \left[ W_2(t) \,\sum\limits_{n \in {\mathbb Z}} x[n] \, \delta(t - n/B) \right],
\end{eqnarray}\normalsize}where $\star$ denotes convolution. $W_2(t)$ limits $x(t)$ to the TD interval $[-(\tau_{\max}+T)/2 \,,\, (\tau_{\max} + T)/2]$ and $w_1(t)$ limits the signal to the FD interval $[-B/2 - \nu_{\max} \,,\, B/2 + \nu_{\max}]$. As an example, Gaussian pulse shaping with

{\vspace{-4mm}
\small
\begin{align}
    \label{eq:pulse2}
    w_{1}(t) &= \bigg(\frac{2\alpha_{g}(B + 2 \nu_{\max})^2}{\pi}\bigg)^{\frac{1}{4}} e^{-\alpha_{g}(B+ 2 \nu_{\max})^2 t^2}, \nonumber \\
    W_{2}(t) &= \bigg(\frac{2}{\pi \beta_{g}(T+\tau_{\max})^2}\bigg)^{\frac{1}{4}} e^{- \frac{1}{\beta_g} \left(\frac{t}{(T + \tau_{\max})} \right)^2}.
\end{align}\normalsize}
$\alpha_{g} = \beta_{g} = 1.584$ ensures that $99\%$ energy of $w_1(t)$
and $W_2(t)$ is within the frame's FD and TD interval respectively.

In Fig.~\ref{fig1}, IDZT based signal processing is equivalent to that based on the continuous inverse Zak transform. The equivalence follows from two facts, first that

{\vspace{-4mm}
\small
\begin{eqnarray}
\label{eqn144}
    x_{dd}(\tau, \nu) & = & \sum\limits_{k,l \in {\mathbb Z}} x_{dd}[k,l] \delta(\tau - k/B) \, \delta(\nu - l/T)
\end{eqnarray}\normalsize}is the DD domain representation of ${\Tilde x}(t)$. To see this, the TD realization of $x_{dd}(\tau, \nu)$ is given by its continuous inverse Zak transform (with delay and Doppler period parameters 
 $\tau_p = M/B$ and $\nu_p = N/ T = 1/\tau_p$ respectively) (see \cite{zakotfs1, zakotfs2} and Chapter $2$ in \cite{otfsbook})

\begin{figure*}[h]%[t]
\vspace{-9mm}
\centering
\includegraphics[width=15.7cm, height=6.3cm]{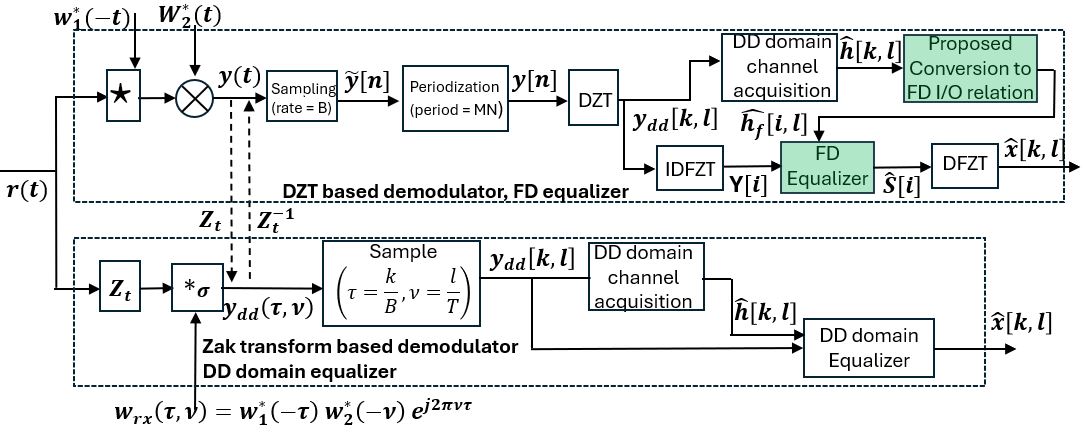}
\vspace{-3mm}
\caption{
Two equivalent implementations of signal processing at the receiver, (i) DZT based processing, and (ii) continuous Zak transform (${\mathcal Z}_t$) based processing.}
\label{fig2}
\vspace{-4mm}
\end{figure*}

{\vspace{-4mm}
\small
\begin{eqnarray}
    {\mathcal Z}_t^{-1}\left( x_{dd}(\tau, \nu)\right) & \mya & \sqrt{\tau_p} \int\limits_{0}^{\nu_p} x_{dd}(t, \nu) \, d\nu \nonumber \\
    & & \hspace{-24mm} \myb \sqrt{\tau_p} \sum\limits_{k \in {\mathbb Z}} \sum\limits_{l \in {\mathbb Z}} x_{dd}[k,l]  \delta(t - k/B) \int\limits_{0}^{\nu_p} \, \delta\left(\nu - l \frac{\nu_p}{N} \right)  d\nu \nonumber \\
    & & \hspace{-24mm} \myc \sqrt{\tau_p} \sum\limits_{k \in {\mathbb Z}} \underbrace{\sum\limits_{l=0}^{N-1} x_{dd}[k,l] }_{= \sqrt{N} x[k] } \delta(t - k/B)  \, = \, {\Tilde x}(t),
\end{eqnarray}\normalsize}where step (a) follows from the definition of the inverse Zak transform, step (b) follows from (\ref{eqn144}) and step (c) follows (\ref{eqn66n}). Second,
pulse shaping of ${\Tilde x}(t)$ into $x(t)$ can be equivalently performed in the DD domain by twisted convolution of $x_{dd}(\tau, \nu)$ with the pulse shaping filter $w_{tx}(\tau, \nu)$ (see Fig.~\ref{fig1})

{\vspace{-4mm}
\small
\begin{eqnarray}
\label{wtxfilt}
    w_{tx}(\tau, \nu) \,  \Define \,  w_1(\tau) \, w_2(\nu) \,,\,
    w_2(\nu)  \Define  \int W_2(t) \, e^{-j 2 \pi \nu t} \, dt.
\end{eqnarray}\normalsize}To see this, note that (\ref{pulseshapeeqn}) can be equivalently written as

{\vspace{-4mm}
\small
\begin{eqnarray}
\label{xteqn15}
    x(t) & = & 
    %\int w_1(\tau) W_2(t - \tau) {\Tilde x}(t - \tau) \, d\tau \nonumber \\
    %& & \hspace{-10mm} = \iint w_1(\tau) w_2(\nu) {\Tilde x}(t - \tau) \, e^{j 2 \pi \nu (t - \tau)} \, d\tau \, d\nu  \nonumber \\
   % & & \hspace{-10mm} = 
    \iint w_{tx}(\tau, \nu) {\Tilde x}(t - \tau) \, e^{j 2 \pi \nu (t - \tau)} \, d\tau \, d\nu.
\end{eqnarray}\normalsize}The DD domain realization of $x(t)$ is given by its continuous Zak transform and is simply (see Chapter $2$ in \cite{otfsbook}, for a similar derivation, starting from the expression in the RHS of (\ref{xteqn15}))

{\vspace{-4mm}
\small
\begin{eqnarray}
{\mathcal Z}_t \left( x(t) \right) & = & x_{dd}(\tau, \nu) \, *_{\sigma} \, w_{tx}(\tau, \nu).
\end{eqnarray}\normalsize}where $*_{\sigma}$ denotes the twisted convolution operation \cite{zakotfs1, zakotfs2, otfsbook}.

With i.i.d. information symbols $x[k,l]$ having zero mean and variance $E$, unit energy filters (i.e., $\int \vert w_1(t) \vert^2 dt = \int \vert W_2(t) \vert^2 dt = 1$), it follows that the average power of $x(t)$ is ${\mathbb E}\left[ \int \vert x(t) \vert^2 \, dt \right]/(T + \tau_{\max}) \approx B T E / (T + \tau_{\max})$. With normalized channel path gains such that the average received and transmit power are same, the received signal to noise ratio (SNR) is
$\rho = \frac{E/N_0}{(1 + \frac{\tau_{\max}}{T}) (1 + \frac{2 \nu_{\max}}{B})}$ since the AWGN power at the receiver is $(B + 2 \nu_{\max}) N_0$.

\subsection{Receiver signal processing}
In the Discrete Zak Transform (DZT) based receiver processing (see top chain in Fig.~\ref{fig2}), the received signal $r(t)$ is match filtered to give 

{\vspace{-4mm}
\small
\begin{eqnarray}
\label{yteqn12}
y(t) & = & \frac{1}{\sqrt{T}} W_2^*(t) \, \left[ w_1^*(-t) \, \star \, r(t) \right].
\end{eqnarray}\normalsize}Sampling at $t = n/B, n \in {\mathbb Z}$ gives the DT signal

{\vspace{-4mm}
\small
\begin{eqnarray}
\label{tildeyneqn}
    {\Tilde y}[n] & = & y\left(t = n/B \right) \,,\, n \in {\mathbb Z}.
\end{eqnarray}\normalsize}Periodization with period $MN$ gives the DT periodic signal

{\vspace{-4mm}
\small
\begin{eqnarray}
\label{periodyneqn}
    y[n] & = & \sum\limits_{p \in {\mathbb Z}} {\Tilde y}[n + p MN] \,,\, n \in {\mathbb Z}.
\end{eqnarray}\normalsize}

%Discrete Frequency Zak transform (DFZT) of $Y[i]$ gives its discrete DD realization
%
%{\vspace{-4mm}
%\small
%\begin{eqnarray}
%y_{dd}[k,l] & = & \frac{1}{\sqrt{M}} \sum\limits_{p=0}^{M-1} Y[l + pN] \, e^{j 2 \pi (l + pN) \frac{k}{MN}}.
%\end{eqnarray}\normalsize}Note that the composition of DFT and DFZT is referred to as the Discrete Zak Transform (DZT) \cite{otfsbook}
%which converts $y[n]$ directly to $y_{dd}[k,l]$

DZT of $y[n]$ gives its discrete DD realization 

{\vspace{-4mm}
\small
\begin{eqnarray}
\label{dzteqn}
    y_{dd}[k,l] & \hspace{-3mm} = & \hspace{-3mm} \mbox{\small{DZT}}(y[n]) =  \hspace{-1mm} \sum\limits_{q=0}^{N-1} \hspace{-1mm} y[k + q M] \, \frac{e^{-j 2 \pi q l/N}}{\sqrt{N}},\, k,l \in {\mathbb Z}.
\end{eqnarray}\normalsize}
The signal processing chain at the bottom of Fig.~\ref{fig2} based on the continuous Zak transform is equivalent to DZT based processing in the top chain, since matched filtering can be equivalently performed in the DD domain, through twisted convolution of the DD representation of $r(t)$ with the DD domain filter $w_{rx}(\tau, \nu) = w_1^*(-\tau) \, w_2^*(-\nu) \, e^{j 2 \pi \nu \tau}$. To see this, note that

{\vspace{-4mm}
\small
\begin{eqnarray}
    y(t) & \hspace{-2.5mm} = & \hspace{-2.5mm} 
    %\frac{1}{\sqrt{T}} W_2^*(t) \left[ w_1^*(-t) \star r(t)\right] \nonumber \\
   % & & \hspace{-20mm} = \frac{1}{\sqrt{T}} \left( \int w_2^*(-\nu) e^{j 2 \pi \nu t} \, d\nu \right) \int w_1^*(-\tau) r(t - \tau) \, d\tau \nonumber \\
  %  & & \hspace{-20mm} = \frac{1}{\sqrt{T}} \iint w_1^*(-\tau) w_2^*(-\nu) e^{j 2 \pi \nu \tau} \, r(t - \tau) \, e^{j 2 \pi \nu (t - \tau)} \, d\tau \, d\nu \nonumber \\
     %& & \hspace{-20mm} = 
     \frac{1}{\sqrt{T}} \iint w_{rx}(\tau, \nu) \, r(t - \tau) \, e^{j 2 \pi \nu (t - \tau)} \, d\tau \, d\nu.
\end{eqnarray}\normalsize}The continuous Zak transform for $y(t)$ then gives

{\vspace{-4mm}
\small
\begin{eqnarray}
\label{eqn98326654}
    y_{dd}(\tau, \nu) & \hspace{-2.5mm} \Define &  \hspace{-2.5mm} {\mathcal Z}_t(y(t)) \, = \, \frac{1}{\sqrt{T}} w_{rx}(\tau, \nu) \, *_{\sigma} \, {\mathcal Z}_t(r(t)).
\end{eqnarray}\normalsize}For equivalence of the two signal processing chains in Fig.~\ref{fig2}, it remains to show that
$y_{dd}(\tau, \nu)$ sampled on $(k/B, l/T), k,l \in {\mathbb Z}$ gives $y_{dd}[k,l]$. Indeed
%which is the discrete DD domain realization of $y[n]$
%(see (\ref{tildeyneqn}), (\ref{periodyneqn}), (\ref{dzteqn}) and (\ref{eqn98326654})). Indeed

{\vspace{-4mm}
\small
\begin{eqnarray}
    y_{dd}\left(\tau = \frac{k}{B} \,,\, \nu = \frac{l}{T} \right) & \hspace{-3mm} \mya & \hspace{-3mm} \sqrt{\tau_p} \sum\limits_{q' \in {\mathbb Z}} y(k/B + q'\tau_p)  \, e^{-j 2 \pi q' l \frac{\tau_p}{T}} \nonumber \\
    & & \hspace{-27mm} \myb \sqrt{T} \sum\limits_{q' \in {\mathbb Z}} \underbrace{y\left( (k + q'M) \frac{\tau_p}{M}\right)}_{= {\Tilde y}[k + q'M]} \, e^{-j 2 \pi q' \frac{l}{N}} \nonumber \\
     & & \hspace{-27mm} \myc \sqrt{T} \sum\limits_{q=0}^{N-1} \underbrace{\sum\limits_{p \in {\mathbb Z}} {\Tilde y}[k + qM + pMN]}_{= y[k + q M]} \, e^{-j 2 \pi q' \frac{l}{N}} \nonumber \\
      & & \hspace{-27mm} \myd \sqrt{T} \sum\limits_{q=0}^{N-1} {y}[k + qM] \, e^{-j 2 \pi q \frac{l}{N}} \, = \, \sqrt{T} y_{dd}[k,l],
\end{eqnarray}\normalsize}where step (a) follows from the definition of ${\mathcal Z}_t$, step (b) follows from $1/B = \tau_p/M$ and $\tau_p/T = 1/N$. Step (c) and (d) follow from (\ref{periodyneqn}) and the definition of DZT in (\ref{dzteqn}).

\subsection{DD domain I/O relation}
Since the two signal processing chains at the transmitter and at the receiver are equivalent, the DD domain I/O relation with IDZT at the transmitter and DZT at the receiver (top chains in Fig.~\ref{fig1} and Fig.~\ref{fig2}) is same as that with continuous DD domain processing (bottom chains in Fig.~\ref{fig1} and Fig.~\ref{fig2}), which is given by \cite{otfsbook}

{\vspace{-4mm}
\small
\begin{eqnarray}
\label{ddiorel}
y_{dd}[k,l] =
%h_{dd}[k,l] \, *_{\sigma} \, x_{dd}[k,l] \, + \, n_{dd}[k,l] \nonumber \\
%&  &  \hspace{-22mm} =  \hspace{-2.75mm} 
\hspace{-3mm} \sum\limits_{k', l' \in {\mathbb Z}} \hspace{-2.5mm} h_{dd}[k',l']x_{dd}[k - k', l- l'] e^{j 2 \pi l' \frac{(k - k')}{MN}}\hspace{-1mm} +\hspace{-1mm}  n_{dd}[k,l],
\end{eqnarray}\normalsize}where $n_{dd}[k,l]$ is the filtered and sampled AWGN. Also

{\vspace{-4mm}
\small
\begin{eqnarray}
    h_{dd}[k,l] & \hspace{-3mm}  = &  \hspace{-3mm} h_{dd}\left(\tau = \frac{k}{B} \,,\, \nu = \frac{l}{T} \right), \nonumber \\
    h_{dd}(\tau, \nu) &  \hspace{-3mm} = & \hspace{-3mm} w_{rx}(\tau, \nu) \, *_{\sigma} \, h_{\mbox{\scriptsize{phy}}}(\tau, \nu) \, *_{\sigma} \, w_{tx}(\tau, \nu).
\end{eqnarray}\normalsize}Due to quasi-periodicity of $x_{dd}[k,l]$, (\ref{ddiorel}) can be equivalently written as an $MN$-periodic discrete twisted convolution with the $MN$-periodic extension of $h_{dd}[k,l]$ (see \cite{spreadpaper}), i.e.

{\vspace{-4mm}
\small
\begin{eqnarray}
\label{eqn8524}
    y_{dd}[k,l] & = & h[k,l] \, \circledast_{\sigma} \, x_{dd}[k,l] \, + \, n_{dd}[k,l], \nonumber \\
    & & \hspace{-22mm} = \hspace{-3mm} \sum\limits_{k'=0}^{MN-1} \sum\limits_{l'=0}^{MN-1} \hspace{-2.5mm} h[k',l]'\, x_{dd}[k - k', l- l']e^{j 2 \pi l' \frac{(k - k')}{MN}}\hspace{-1mm}+ n_{dd}[k,l],
\end{eqnarray}\normalsize}where the $MN$-periodic extension $h[k,l]$ is given by
\begin{eqnarray}
    h[k,l] & = & \sum\limits_{n,m \in {\mathbb Z}} h_{dd}[k + nMN, l + mMN].
\end{eqnarray}The effective channel $h_{dd}[k,l]$ can be acquired/estimated efficiently as long as the crystallization condition is satisfied, i.e., $M$ and $N$ are greater than
the spreads of $h_{dd}[k,l]$ along the discrete delay and Doppler axis, respectively \cite{zakotfs2, otfsbook}. The I/O relation in (\ref{eqn8524}) can be expressed in the matrix vector form as in \cite{zakotfs2, otfsbook}. The effective $MN \times MN$ DD domain channel matrix lacks structure and therefore the complexity for linear equalization is \emph{high} (${\mathcal O}(M^3 N^3)$) (as an $MN \times MN$ matrix has to be inverted). In this paper, we propose equalization in frequency domain (FD) where we show that the effective FD channel matrix is sparse and has a \emph{banded} structure due to which linear equalization has complexity \emph{only} $O(M^2 N^2)$ (see Section \ref{seczakofdm}). FD equalization however requires acquisition of the comlete FD I/O relation. In the next section, we propose a novel low-complexity method to \emph{convert the acquired DD domain I/O relation to FD}.

\section{Converting DD domain I/O relation to FD}
The FD I/O relation is between the FD realizations of the transmitted and the received DD signals. Let $S[i], i \in {\mathbb Z}$ denote the discrete FD realization of the transmit DD signal $x_{dd}[k,l]$ given by its Inverse Discrete Frequency Zak transform (IDFZT) (see Appendix of Chapter $8$ in \cite{otfsbook})

{\vspace{-4mm}
\small
\begin{eqnarray}
    S[i] & = & \frac{1}{\sqrt{M}} \sum\limits_{k=0}^{M-1} x_{dd}[k,i] \, e^{-j 2 \pi \frac{i k}{MN}}.
\end{eqnarray}\normalsize}Note that $S[i] = S[i + pMN]$ for all $p \in {\mathbb Z}$.

%Also, for $p \in {\mathbb Z}$, from the periodicity of $x_{dd}[k,l]$ along the Doppler axis with period $N$, it follows that

%{\vspace{-4mm}
%\small
%\begin{eqnarray}
%\label{eqnddspread}
%S[i+pN] %&  \hspace{-3mm} =  & \hspace{-3mm}  \frac{1}{\sqrt{M}} \sum\limits_{k=0}^{M-1} x_{dd}[k, i + pN]  \, e^{-j 2 \pi \frac{(i + pN) k}{MN}} \nonumber \\
%&  \hspace{-3mm} =  & \hspace{-3mm}   \frac{1}{\sqrt{M}} \sum\limits_{k=0}^{M-1} x_{dd}[k, i]  \, e^{-j 2 \pi \frac{(i + pN) k}{MN}} ,
%\end{eqnarray}\normalsize}i.e., for each $i=0,1,\cdots, N-1$, the $M$ information symbols $x_{dd}[k,i] = x[k,i], k=0,1,\cdots, M-1$ are jointly precoded across the $M$ FD symbols $S[(i + pN) \, \mbox{\small{mod}}  \, MN], p=0,1,\cdots, M-1$.

Similarly, the discrete FD realization of $y_{dd}[k,l]$ is given by its IDFZT

{\vspace{-4mm}
\small
\begin{eqnarray}
    Y[i] & = & \frac{1}{\sqrt{M}} \sum\limits_{k=0}^{M-1} y_{dd}[k,i] \, e^{-j 2 \pi \frac{i k}{MN}}.
\end{eqnarray}\normalsize}Note that $Y[i] = Y[i + pMN]$ for all $p \in {\mathbb Z}$.
The following novel result is the basis for the proposed conversion of the DD domain I/O relation to that in FD.
\label{sec4}
\begin{theorem}
The FD I/O relation between $Y[\cdot]$ and $S[\cdot]$ is

    {\vspace{-4mm}
\small
\begin{eqnarray}
\label{eqn3232}
    Y[i] & \hspace{-2mm} = & \hspace{-2mm} \sum\limits_{k'=0}^{MN-1} \sum\limits_{l'=0}^{MN-1}  h[k', l'] S[i - l'] \, e^{-j 2 \pi \frac{i k'}{MN}}  \, + \, Z[i] \nonumber \\
    & = & \sum\limits_{l'=0}^{MN-1} h_f[i, l'] \, S[l']  \, + \, Z[i],
\end{eqnarray}\normalsize}where $Z[i] = \sum\limits_{k=0}^{M-1} n_{dd}[k,i]/\sqrt{M}$ is the AWGN in the $i$-th subcarrier. Further, $h_f[i,l]$ describes the FD I/O relation completely and is \emph{computed} from the effective discrete DD channel taps $h[k,l]$ using (see green shaded block in Fig.~\ref{fig2})

{\vspace{-4mm}
\small
\begin{eqnarray}
\label{eqn3332}
    h_f[i,l] & = & \sum\limits_{k'=0}^{MN-1}  h[k', (i - l)]  \, e^{-j 2 \pi \frac{i k'}{MN}}.
\end{eqnarray}\normalsize}
\end{theorem}
\begin{IEEEproof}
Taking IDFZT of both sides of (\ref{eqn8524}) gives (\ref{eqn2828}) (see top of this page).
\begin{figure*}
\vspace{-9mm}
{\small
\begin{eqnarray}
\label{eqn2828}
Y[i] & \hspace{-3mm} \mya & \hspace{-3mm} \frac{1}{\sqrt{M}} \hspace{-2mm} \sum\limits_{k=0}^{M-1} \hspace{-1.5mm} y_{dd}[k,i] \, e^{-j 2 \pi \frac{i k}{MN}}
 \myb \hspace{-2mm} \sum\limits_{k'=0}^{MN-1} \sum\limits_{l'= 0}^{MN-1} \hspace{-3mm} h[k',l'] e^{-j 2 \pi \frac{i k'}{MN}}  \left( \frac{1}{\sqrt{M}} \sum\limits_{k=0}^{M-1} \hspace{-1.5mm} x_{dd}[k - k', i - l'] \, e^{-j 2 \pi \frac{(k - k') (i - l')}{MN}}\right) + Z[i].
\end{eqnarray}\normalsize}
\vspace{-4mm}
\begin{eqnarray*}
\hline
\end{eqnarray*}
\vspace{-11mm}
\end{figure*}
In (\ref{eqn2828}), step (a) follows from the fact that $Y[i]$ is the discrete FD representation of $y_{dd}[k,l]$. Step (b) follows from substituting $y_{dd}[k,l]$ from (\ref{eqn8524}) into the RHS of step (a). Observe that $x_{dd}[k - k', i - l'] \, e^{-j 2 \pi \frac{(k - k') (i - l')}{MN}}$ is periodic in $k$ with period $M$ since for any $n \in {\mathbb Z}$
\begin{eqnarray}
    x_{dd}[k - k'+ nM, i -l'] e^{-j 2 \pi  \frac{(k - k' + nM) (i - l')}{MN} } & & \nonumber \\
    & & \hspace{-69mm} \mya x_{dd}[k - k', i - l'] \, e^{j 2 \pi n \frac{(i - l') }{N}} \, e^{-j 2 \pi  \frac{(k - k' + nM) (i - l')}{MN} } \nonumber \\
    & & \hspace{-69mm} = x_{dd}[k - k', i - l'] \, e^{-j 2 \pi \frac{(k - k') (i - l')}{MN}},
\end{eqnarray}where step (a) follows from the quasi-periodicity of $x_{dd}[k,l]$. Therefore the terms within the brackets in the RHS of (\ref{eqn2828}) is

{\vspace{-4mm}
\small
\begin{eqnarray}
    \frac{1}{\sqrt{M}} \sum\limits_{k=0}^{M-1} x_{dd}[(k - k') \, mod \, M, i - l'] \, e^{-j 2 \pi \frac{(i - l') ((k - k') \, mod \, M)}{MN}} & & \nonumber \\
    & & \hspace{-90mm} = \frac{1}{\sqrt{M}} \sum\limits_{k=0}^{M-1} x_{dd}[k, i - l'] \, e^{-j 2 \pi \frac{k (i - l') }{MN}}  \, = \, S[i - l']  
\end{eqnarray}\normalsize}since the discrete FD representation of $x_{dd}[k,l]$ is $S[i]$. Using this in (\ref{eqn2828}) completes the proof,
where the second step in (\ref{eqn3232}) follows from the $MN$-periodicity of $S[\cdot]$ and that $h[k,l]$ is also $MN$-periodic along both axes.
\end{IEEEproof}

Since $h_{dd}(\tau, \nu) = w_{rx}(\tau, \nu) *_{\sigma} h_{\mathrm{phy}}(\tau, \nu) *_{\sigma} w_{tx}(\tau, \nu)$ takes significant values only for $\vert \nu \vert < \mathcal{O}(\nu_{\max})$, for any $k$, $h[k,l]$ will have significant values only for $\vert l \vert \leq l_{\max}$, where $l_{\max} = \mathcal{O}(\lceil T \nu_{\max} \rceil)$. Therefore, $h_f[i,l']$ in (\ref{eqn3232}) has significant values only for $\vert i - l' \vert \leq l_{\max}$. {Given $h[k,l]$, the complexity of computing $h_f[i,l]$ in (\ref{eqn3332}) is therefore ${\mathcal O}((2l_{\max} + 1) M^2N^2)$}.

The FD I/O relation in (\ref{eqn3232}) has the matrix-vector form as in (\ref{eqn3333}) (see top of next page).
\begin{figure*}
\vspace{-9mm}
{\small
\begin{eqnarray}
\label{eqn3333}
{\bf Y} & = & {\bf H} \, {\bf S} + {\bf Z}, \,\, \,
{\bf Y} \, \Define \, \left[ Y[0] , Y[1], \cdots, Y[MN-1] \right]^T, \,\, 
{\bf Z} \, \Define \, \left[ Z[0] , Z[1], \cdots, Z[MN-1]\right] \nonumber \\
{\bf S} & \hspace{-3mm} \Define & \hspace{-3mm} \left[  S[MN - l_{\max}], \cdots, S[MN - 1] , S[0], \cdots, S[MN-1], S[0], \cdots, S[l_{\max} - 1] \right]^T.
\end{eqnarray}\normalsize}
\vspace{-7mm}
\begin{eqnarray*}
\hline
\end{eqnarray*}
\vspace{-13mm}
\end{figure*}
The effective FD channel matrix ${\bf H} \in {\mathbb C}^{MN \times (MN + 2 l_{\max})}$ is \emph{banded}. Let $H[i,l']$ denote the element of ${\bf H}$ in its $i$-th row and $l'$-th column, $i=0,1,\cdots, MN-1$, $l'= 0, 1, \cdots, MN + 2l_{\max} - 1$, then

{\vspace{-4mm}
\small
\begin{eqnarray}
\label{Hmateqn}
    H[i , i+ l + l_{\max}] & \hspace{-3mm} = \begin{cases}
        h_f[i, -l] &, i=0,1,\cdots, MN-1, \\
        & \,\,  \mbox{\small{and}} \, \vert l \vert \leq l_{\max} \\
        0 &, \mbox{\small{otherwise}} \\
    \end{cases}.
\end{eqnarray}\normalsize}
{ 
Note that the I/O relation in (\ref{eqn3232}) corresponds to a type of FD modulation, where
there are $MN$ carriers with carrier spacing equal to $B/(MN) = 1/T$.
%does \emph{not} correspond to OFDM, since the relation between the FD symbols $S[i], i=0,1,\cdots, MN-1$ and the transmit signal $x(t)$ is not that of OFDM.
%i.e., $x(t) \ne \sum\limits_{i=0}^{MN-1} S[i] \, e^{j 2 \pi i t/T}$.
%It however  
}
 
\section{Proposed low-complexity FD equalization of Zak-OTFS}
\label{seczakofdm}
At the receiver (top chain in Fig.~\ref{fig2}), $y_{dd}[k,l]$ is converted to FD realization $Y[i]$ using IDFZT (complexity ${\mathcal O}(MN \log(M))$).
A linear estimate of the vector ${\bf S}$ of transmitted FD symbols is obtained from the FD matrix-vector I/O relation in (\ref{eqn3333}) as (see FD equalizer in Fig.~\ref{fig2})

{\vspace{-4mm}
\small
\begin{eqnarray}
\label{eqnshat}
{\widehat {\bf S}} & \Define & {\bf H}^H \, \left({\bf H} {\bf H}^H + {{\bf I}}/{\rho} \right)^{-1} \, {\bf Y}
\end{eqnarray}\normalsize}whose elements are the estimated FD symbols, i.e., $\widehat{S}[i], i=0,1,\cdots, MN-1$. An estimate of the information symbols is then given by the DFZT of $\widehat{S}[i]$ restricted over one period along delay and Doppler, i.e.

{\vspace{-4mm}
\small
\begin{eqnarray}
\widehat{x}[k,l] & \Define &  \frac{1}{\sqrt{M}} \sum\limits_{p=0}^{M-1} \widehat{S}[l+ pN] \, e^{j 2 \pi (l + pN) \frac{k}{MN}},
\end{eqnarray}\normalsize}$k=0,1,\cdots, M-1$, $l=0,1,\cdots , N-1$. These are then rounded off to the nearest modulation symbol. The complexity of DFZT is ${\mathcal O}(MN \log(M))$. Since ${\bf H}$ is banded, $\left({\bf H} {\bf H}^H + {{\bf I}}/{\rho} \right)$ is also banded with width of the band $b = (4 l_{\max} + 1)$. Banded structure allows for low-complexity inversion in (\ref{eqnshat})  with complexity only ${\mathcal O}(b \, M^2 N^2)$ (see \cite{Bandedinverse2}) as compared to the ${\mathcal O}(M^3N^3)$ complexity of inversion of the DD domain channel matrix.
In practice, there is always some small energy in the matrix elements outside the band, and therefore in the estimated ${\bf H}$, we force elements outside the band to be zero.
Since our focus in this paper is on low-complexity FD equalization of Zak-OTFS, we assume perfect knowledge of the DD domain channel $h[k,l]$ from which ${\bf H}$ is derived using (\ref{eqn3332}) and (\ref{Hmateqn}). From previous results in \cite{zakotfs2, otfsbook}, it is known that $h[k,l]$ is accurately estimated by transmitting a pilot in the DD domain along with information symbols. Note that we can avoid pilot overhead by using a spread pilot as described in \cite{spreadpaper}. The complexity of estimating $h[k,l]$ at the receiver is ${\mathcal O}(k_{\max}l_{\max}MN)$ where $k_{\max}$ is the delay spread of $h_{dd}[k,l]$.

\section{Numerical results}
\label{simsec}
\begin{figure}[!t]
    \hspace{-2mm}
    \includegraphics[width=9.6cm, height=7.0cm]{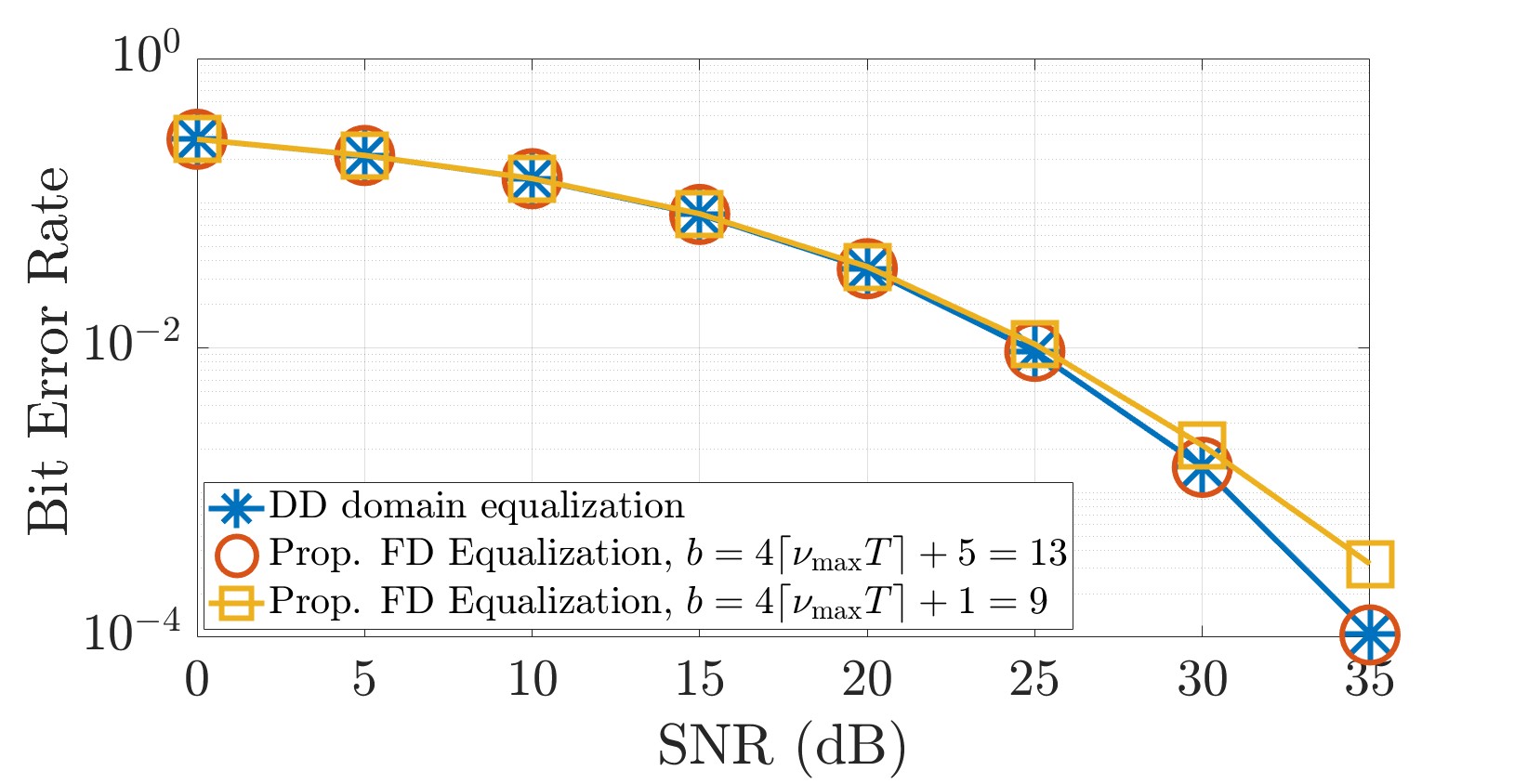}
    \vspace{-8mm}
    \caption{Uncoded $4$-QAM BER vs SNR.}
    \label{fig3}
    \vspace{-5mm}
\end{figure}

In Fig.~\ref{fig3} we plot the uncoded $4$-QAM bit error rate (BER) vs. SNR ($\rho$)
 for the Veh-A channel in \cite{EVAITU} (Table~\ref{tab:veh_a}). In the Veh-A channel model, $h_{\mbox{\scriptsize{phy}}}(\tau, \nu) = \sum\limits_{i=1}^6 h_i \delta(\tau - \tau_i) \delta(\nu - \nu_i)$, where $h_i, \tau_i, \nu_i$ are the complex gain, delay and Doppler shift of the $i$-th path. $\nu_i = \nu_{max} \cos(\theta_i)$, where $\theta_i, i=1,2,\cdots, 6$ are i.i.d. uniformly distributed in $[0 \,,\, 2 \pi)$. Complex gains $h_i, i=1,2,\cdots, 6$ are independent complex Gaussian distributed with the relative power in Table~\ref{tab:veh_a} equal to ${\mathbb E}[\vert h_i \vert^2]/{\mathbb E}[\vert h_1 \vert^2]$. We use a Gaussian pulse shaping filter as in (\ref{eq:pulse2}). The other parameters are, $M=31, N=37$, Doppler period $\nu_p = 30$ KHz, $T = 1.23$ ms, $B = 930$ kHz, $\tau_{\max} = 2.51 \mu$s, $\nu_{\max} = 815$ Hz.

With a Gaussian filter, $l_{\max} = 1 + \lceil T \nu_{\max} \rceil$, i.e., $b = 4 \lceil T \nu_{\max} \rceil + 5$, for which the proposed FD equalization (with the proposed I/O relation conversion from DD domain to FD) achieves same performance as DD domain equalization (with DD domain channel acquisition) but with a significantly lower complexity of $\mathcal{O}(bM^2N^2)$ compared to $\mathcal{O}(M^3N^3)$ for DD domain equalization. To reduce the complexity further, we can truncate the band so that its width is smaller, but then this results in slight performance degradation at high SNR (see curve corresponding to $b = 4 \lceil T \nu_{\max} \rceil + 1$). 

\begin{table}[!t]
    \centering
    \caption{Power-delay profile of Veh-A channel model}
    \vspace{-2mm}
    \begin{tabular}{|c|c|c|c|c|c|c|}
         \hline
         Path index $i$ & 1 & 2 & 3 & 4 & 5 & 6 \\
         \hline
         Delay $\tau_i (\mu s)$ & 0 & 0.31 & 0.71 & 1.09 & 1.73 & 2.51 \\
         \hline
         Relative power (dB) & 0 & -1 & -9 & -10 & -15 & -20 \\
         \hline
    \end{tabular}
    \label{tab:veh_a}
    \vspace{-3mm}
\end{table}

%\begin{figure}[!t]
%    \centering
%    \includegraphics[width=7.0cm, height=4.9cm]{channel_fading_tf_vs_dd.jpg}
%    \vspace{-3mm}
%    \caption{Ratio of the received energy (integrated over all carriers) of a signal transmitted on a particular carrier to the energy transmitted on that carrier.}
%    \label{fig4}
%    \vspace{-5mm}
%\end{figure}

\section{Conclusion}
We have introduced a method of equalizing Zak-OTFS modulation in the FD that takes advantage of the banded structure of the effective FD channel matrix.
Our method reduces the complexity of equalization from ${\mathcal O}(M^3 N^3)$
to ${\mathcal O}(M^2 N^2)$ where $MN$ is the number of Zak-OTFS carriers. We have also proposed a novel low-complexity method for constructing the FD effective channel from the acquired DD domain channel. For simplicity, we assume perfect channel state information at the receiver, deferring channel estimation, and other practical considerations to future work.

\section{Acknowledgement}
The authors thank Ronny Hadani for many valuable discussions.
%\textcolor{blue}{The authors thank Ronny Hadani for pointing out that the
%FD I/O relation in (\ref{eqn3232}) does not correspond to OFDM.}


\begin{thebibliography}{1}
\bibitem{Nee2000}
R. V. Nee, and R. Prasad, “OFDM for Wireless Multimedia Communications,” Artech House Inc., 2000.

\bibitem{Wang2006}
T. Wang, J. G. Proakis, E. Masry and J. R. Zeidler, “Performance degradation of OFDM systems due to Doppler spreading,” \emph{IEEE Trans. on Wireless Commun.}, vol. 5, no. 6, June 2006.

\bibitem{zakotfs1}
S. K. Mohammed, R. Hadani, A. Chockalingam and R. Calderbank, “OTFS - A mathematical foundation for communication and radar sensing in the delay-Doppler domain," \emph{IEEE BITS the Information Theory Magazine}, vol. 2, no. 2, pp. 36–55, 1 Nov. 2022.

\bibitem{zakotfs2}
S. K. Mohammed, R. Hadani, A. Chockalingam and R. Calderbank, ``OTFS—Predictability in the Delay-Doppler Domain and Its Value to Communication and Radar Sensing," \emph{IEEE BITS the Information Theory Magazine}, vol. 3, no. 2, pp. 7-31, June 2023.

\bibitem{otfsbook}
{S. K. Mohammed, R. Hadani and A. Chockalingam, ``OTFS Modulation: Theory and Applications," \emph{IEEE Press and Wiley}, Nov. 2024.}

\bibitem{spreadpaper}
M. Ubadah, S. K. Mohammed, R. Hadani, S. Kons, A. Chockalingam and R. Calderbank, ``Zak-OTFS to Integrate Sensing the I/O Relation and Data Communication," submitted to IEEE, arXiv:2404.04182v2.

\bibitem{Hadani2017}
R. Hadani, S. Rakib, M. Tsatsanis, A. Monk, A. J. Goldsmith, A. F. Molisch, and R. Calderbank, “Orthogonal time frequency space modulation,” \emph{Proc. IEEE WCNC’2017}, pp. 1-6, Mar. 2017.


        \bibitem{Bello}
        P.~A.~Bello, ``Characterization of Randomly Time-Variant Linear Channels," {\em IEEE Trans. Comm. Syst.}, vol. 11, pp. 360-393, 1963. 

%\bibitem{Bandedinverse}
%R. S. Ran and T. Z. Huang, ``An inversion algorithm for a banded matrix," \emph{Computers \& Mathematics with Applications}, vol. 58, no. 9, pp. 1699 - 1710, Nov. 2009.

\bibitem{Bandedinverse2}
A. Mahmood, D. J. Lynch and L. D. Philipp, ``A fast banded matrix inversion using connectivity of Schur's complements," \emph{IEEE 1991 Int. Conf. on Systems Engineering}, Dayton, OH, USA, 1991, pp. 303-306.



\bibitem{EVAITU} 
ITU-R M.1225, ``Guidelines for evaluation of radio transmission technologies for IMT-2000,'' {\it Int. Telecom. Union Radio communication}, 1997. 
	


\bibitem{Hanly23}
S. Gopalam, I. B. Collings, S. V. Hanly, H. Inaltekin, S. R. B. Pillai and P. Whiting, ``Zak-OTFS Implementation via Time and Frequency Windowing," \emph{IEEE Trans. on Communications}, vol. 72, no. 7, July 2024.


	\end{thebibliography}
\end{document}